# Microstructure engineering of metamagnetic Ni-Mn-based Heusler compounds by Fe-doping: A roadmap towards excellent cyclic stability combined with large elastocaloric and magnetocaloric effects

Lukas Pfeuffer [a,*], Jonas Lemke [a], Navid Shayanfar [a], Stefan Riegg [a], David Koch [b], Andreas Taubel [a], Franziska Scheibel [a], Nagaarjhuna A. Kani [a,c], Esmaeil Adabifiroozjaei [c], Leopoldo Molina-Luna [c], Konstantin P. Skokov [a] and Oliver Gutfleisch [a]

[a] Functional Materials, Institute of Materials Science, Technical University of Darmstadt, 64287 Darmstadt, Germany
[b] Structure Research, Institute of Materials Science, Technical University of Darmstadt, 64287 Darmstadt, Germany
[c] Advanced Electron Microscopy, Institute of Materials Science, Technical University of Darmstadt, 64287 Darmstadt, Germany

**Abstract**

Ni-Mn-based metamagnetic shape-memory alloys exhibit a giant thermal response to magnetic fields and uniaxial stress which can be utilized in single caloric or multicaloric cooling concepts for energy-efficient and sustainable refrigeration. However, during cyclic operation these alloys suffer from structural and functional fatigue as a result of their high intrinsic brittleness. Here, we present based on Fe-doping of Ni-Mn-In a microstructure design strategy which simultaneously improves cyclic stability and maintains the excellent magnetocaloric and elastocaloric properties. Our results reveal that precipitation of a strongly Fe-enriched and In-depleted coherent secondary γ-phase at grain boundaries can ensure excellent mechanical stability by hindering intergranular fracture during cyclic loading. In this way, a large elastocaloric effect of -4.5 K was achieved for more than 16000 cycles without structural or functional degradation, which corresponds to an increase of the cyclic stability by more than three orders of magnitude as compared to single-phase Ni-Mn-In-(Fe). In addition, we demonstrate that the large magnetocaloric effect of single-phase Ni-Mn-In-(Fe) can be preserved in the dual-phase material when the secondary γ-phase is exclusively formed at grain boundaries as the martensitic transformation within the Heusler matrix is barely affected. This way, an adiabatic temperature change of -3 K and an isothermal entropy change of 15 Jkg$^{-1}$K$^{-1}$ was obtained in 2 T for dual-phase Ni-Mn-In-Fe. We expect that this concept can be applied to other single caloric and mutlicaloric materials, therewith paving the way for solid-state caloric cooling applications.

## 1 Introduction

Refrigeration, including space cooling, accounts already for about 20 % of the global electricity consumption. At the same time, the demand for artificial cold is rapidly increasing due to a rising prosperity in many parts of the world [1]. Today, cooling systems are predominantly based on vapor-compression technology having a low thermodynamic efficiency in small-scale applications and a substantial environmental impact [2,3]. On the one hand, this impact is caused by $CO_2$ emissions during electricity generation via fossil fuels to power cooling devices. On the other hand, utilized refrigerants are based on hazardous, ozone-depleting or greenhouse gases [4]. In consequence, there is a high demand for energy-efficient and environmentally-friendly alternatives to vapor-compression technology.

The most promising options are solid-state caloric cooling techniques such as magneto-, electro-, baro- or elastocaloric cooling which utilize the thermal response of a ferroic material when exposed to a magnetic field, electric field, hydrostatic pressure or uniaxial stress [5]. The key quantities to describe the thermal response are the adiabatic temperature change $\Delta T_{ad}$ and the isothermal entropy change $\Delta s_T$ upon application/removal of the external field. Particularly large values of $\Delta T_{ad}$ and $\Delta s_T$ are obtained at first-order phase transitions due to the associated latent heat. These so-called giant caloric effects were discovered in 1997 in the magnetocaloric material $Gd_5Si_2Ge_2$ [6]. Since then, giant magnetocaloric effects have been discovered in numerous materials such as La(Fe,Si)$_{13}$ [7–9], Fe$_2$P-type [10–12] and Ni-Mn-based Heusler [13–15] compounds. Moreover, magnetocaloric cooling is considered to be the most developed solid-state caloric cooling technique [16]. Besides that, especially giant elastocaloric effects have caught attention by very large $\Delta T_{ad}$ values in reasonable uniaxial stresses [17]. Thus, a $|\Delta T_{ad}|$ of more than 35 K has been observed during unloading of Ni-rich Ti-Ni alloys [18].

* Corresponding author:
    E-mail address: lukas.pfeuffer@tu-darmstadt.de (L. Pfeuffer)

Nevertheless, the utilization of giant caloric materials still suffers from several issues. A major issue is the intrinsic thermal hysteresis of first-order phase transitions limiting the cyclic caloric performance as a result of irreversibilities and energy losses [19,20]. In addition, large volume changes and strains upon the phase transformation can cause mechanical failure/fatigue due to stress concentrations.

An efficient way to improve the cyclic caloric performance is the combination of multiple external fields, which is designated as multicaloric effect [21]. For this purpose, a significant coupling of the ferroic order parameters is essential. Especially metamagnetic shape-memory alloys such as Ni-Mn-based Heusler compounds exhibit a strong magnetoelastic coupling upon the martensitic transformation, which is therefore sensitive to magnetic fields and uniaxial stress [22]. While the magnetic field stabilizes the ferromagnetic austenite, uniaxial stress favors the weak-magnetic martensite phase. In recent works, *Gottschall et al.* [23] and *Gràcia et al.* [24] have demonstrated that both, a sequential as well as a simultaneous combination of magnetic fields and uniaxial stress in ternary Ni-Mn-In can enable cyclic caloric effects which outperform their single caloric counterparts.

Structural and functional fatigue of giant caloric materials during cyclic operation is a major issue, in particular when large uniaxial stresses have to be tolerated. Especially Ni-Mn-based Heusler compounds suffer from a high intrinsic brittleness causing mechanical failure after a few cycles only [25]. Several studies demonstrate that mechanical strengthening of these alloys can be realized in practice by precipitation hardening via doping with various elements [26–30], an optimized texture [31], Hall-Patch [32–34] and solid solution [35,36] strengthening. However, the corresponding functional caloric properties are often not reviewed or solely investigated with respect to a single external stimulus. For the latter, chemical and microstructural changes, introduced for mechanical strengthening, can degrade the caloric properties or change the martensitic transformation temperature causing an incomparability of the caloric effects due their strong temperature dependency [37].

This work sets out to engineer a microstructure which significantly improves the fatigue resistance and maintains the excellent caloric response to both magnetic fields and uniaxial stress of metamagnetic Ni-Mn-based Heusler compounds. For this purpose, we have taken the promising multicaloric Ni-Mn-In system as a model and doped it with different amounts of Fe to adjust various microstructures. At the same time, we kept the martensitic transformation at similar temperature by varying the Mn/In ratio which allows a direct comparison of the functional properties. Our results demonstrate that an optimized microstructural design can combine ultralow fatigue and excellent functionality for single caloric and multicaloric cooling using Ni-Mn-based Heusler compounds.

## 2 Experimental details

Nominally composed $Ni_{49.9}Mn_{35.3+x}In_{14.8-x}Fe_y$ samples with y = 0, 2, 4, 6, 8 were prepared by repeatedly arc melting of elemental Ni, Mn, In and Fe. Thereby, x = 0, 0.4, 0.6, 1.4, 2.2 was adjusted in a way that the martensitic transformation is near room temperature in all samples. In order to compensate Mn evaporation during the melting process, an extra amount of 3 wt% Mn was added to the nominal compositions. Subsequently, the arc-molten ingots were suction cast into cylinders of 3 mm diameter and 30 mm height, annealed for 24 h at 900 °C under Ar and quenched in water. An overview of all samples discussed in this study can be found in Table 1.

X-ray diffraction (XRD) at room temperature was executed in a Stoe Stadi P diffractometer using Mo $K_{\alpha 1}$ radiation and a 2θ range from 5 to 50° of 0.01° step size. Temperature-dependent XRD was conducted in a purpose-built diffractometer with Mo Kα radiation, a 2θ range from 7 to 58° and a step size of 0.009°. A detailed description of the diffractometer can be found in [38]. For the measurements in both setups, a small piece of each suction-cast sample was ground into powder of particle sizes < 40 µm. To guarantee the relaxation of the deformation-induced stresses, the powder was annealed in Ar atmosphere for 20 min at 900 °C, followed by quenching in water. Structural analysis was carried out by Rietveld refinement using the FullProf/WinPlotr software [39,40] and JANA2006 [41] in a superspace approach [42] for modulated martensite.

Microstructure and chemical composition were characterized by backscatter electron imaging (BSE) and energy-dispersive x-ray spectroscopy (EDX) in a Tescan Vega3 and Phillips XL30 FEG high-resolution scanning electron microscope (SEM). Nanoscale structural characterizations were done using a 200 kV JEOL JEM 2100-F Transmission Electron Microscope (TEM). For this purpose, a TEM lamella was prepared by cutting a slice of approximately 100 µm thickness of a bulk sample. Subsequent thinning down to an electron transparent lamella was achieved by an ion milling two-step process using a Gatan 691 Precision Ion Polishing System (PIPS). First the angles and ion beam energy were set to 8° and 5.5 keV, respectively, and milling was done until a small hole was observed in the sample center. In the second step, an angle of 2° and an ion beam energy of 2 keV was used in order to remove the damages from the previous step of milling. A Zeiss Axio Imager.D2M optical microscope equipped with a $LN_2$ cryostat was used for temperature-dependent microstructural investigations. Thereby, a heating and cooling rate of 2 $Kmin^{-1}$ was applied.

Magnetic measurements were performed in a Lake Shore vibrating sample magnetometer (VSM) and a Quantum Design physical property measurement system (PPMS-14 T). Isofield curves of magnetization were accomplished with a heating and cooling rate of 2 $Kmin^{-1}$. Isothermal magnetization measurements were executed up to 2 T with a field-application rate of 0.005 $Ts^{-1}$ in temperature steps of 2 K. A discontinuous temperature protocol was chosen to erase effects of the magnetic field-induced transformation history. For this purpose, the sample was heated after field application and removal to the full austenite and cooled to the full martensite state before the measurement temperature was set. The isothermal magnetization data were utilized for the computation of the isothermal entropy change $\Delta s_T(T,H)$.

Adiabatic temperature changes $\Delta T_{ad}$ were determined by a differential T-type thermocouple attached to the sample in a purpose-built device which generates a sinusoidal magnetic field profile of up to 1.9 T [43]. In agreement with the determination of $\Delta s_T(T,H)$, the abovementioned discontinuous measurement protocol was chosen.

Differential scanning calorimetry (DSC) was executed in a Netzsch DSC 404 F1 Pegasus using a heating and cooling rate of 5 $Kmin^{-1}$. The start and finish temperatures of the martensitic and its reverse transformation as well as the Curie temperature of the austenite $T_C^A$ were determined via the double tangent method.

Compressive strength and elastocaloric experiments were conducted in an Instron 5967 30 kN universal testing machine, equipped with a temperature chamber. For all measurements, cylinders of 5 mm height were cut from the annealed suction-cast rods. The strain along the cylinder axis was recorded by a strain gauge extensometer (Instron 2620) attached to the compression platens and the force was detected via a load cell. The compressive strength tests were performed with a strain rate of $3\times10^{-4}$ s$^{-1}$. In the elastocaloric experiments, each sample was initially mechanically trained by five superelastic cycles to ensure a reproducible material response and subsequently measured using a strain rate of $3\times10^{-2}$ s$^{-1}$ for quasi-adiabatic conditions [44]. The adiabatic temperature change $\Delta T_{ad}$ was recorded by a K-type thermocouple attached to the sample surface.

*Table 1: Chemical compositions of annealed $Ni_{49.9}Mn_{35.3+x}In_{14.8-x}Fe_y$ determined from EDX, secondary phase fraction and average grain diameter of the matrix phase.*

| Sample ID | y | x | Composition matrix | Composition secondary phase | Secondary phase (wt%) | Grain diameter (µm) |
|---|---|---|---|---|---|---|
| Fe0 | 0 | 0 | $Ni_{50.6}Mn_{34.9}In_{14.5}$ | | | 110 ± 24 |
| Fe2 | 2 | 0.6 | $Ni_{49.7}Mn_{34.6}In_{13.5}Fe_{2.2}$ | | | 156 ± 74 |
| Fe4 | 4 | 0.8 | $Ni_{48.2}Mn_{34.3}In_{13}Fe_{4.5}$ | | | 822 ± 215 |
| Fe6 | 6 | 1.5 | $Ni_{48}Mn_{34.5}In_{13}Fe_{4.5}$ | $Ni_{42.1}Mn_{32.9}In_{2.4}Fe_{22.6}$ | 5.2 | 58 ± 14 |
| Fe8 | 8 | 2.3 | $Ni_{47.4}Mn_{35}In_{13.1}Fe_{4.5}$ | $Ni_{37.7}Mn_{35.5}In_{1.6}Fe_{25.2}$ | 11.5 | 60 ± 15 |

## 3 Results and discussion

### 3.1. Microstructural and structural analysis

Figure 1 illustrates the SEM-BSE images of the as-cast (exemplarily for Fe0) and annealed (complete series Fe0 to Fe8) suction-cast samples. In the as-cast state, a radially-symmetric microstructure with an average grain diameter of 57±13 µm is found which results from the crystallization of the material opposite to the heat flow in the cylinder-shaped mold. A very similar microstructure including the average grain diameter can be observed for all as-cast samples independent of the Fe-concentration. In contrast, a grain coarsening from Fe0 to Fe4 can be noticed after annealing. While Fe0 shows an average grain diameter of 110±24 µm, there is an increase to 822±215 µm for Fe4. Besides that, the columnarity has entirely disappeared in Fe4. Upon further addition of Fe, a strongly Fe-enriched and In-depleted secondary γ-phase is preferentially precipitated at grain boundaries during annealing which can be seen for Fe6 and Fe8. According to the EDX results shown in Table 1, the secondary γ-phase is formed as soon as the solubility limit of 4.5 at% Fe in the matrix phase is exceeded. The γ-precipitates cause a grain boundary pinning which results in a refinement of the grains in Fe6 and Fe8 to an average diameter of about 60±15 µm. This value is in excellent agreement with the grain diameter of the as-cast samples. It should be emphasized, that the secondary phase fraction in Fe8 (11.5 vol% ≙ 11.5 wt%) is more than doubled compared to Fe6 (5.2 vol% ≙ 5.2 wt%) causing a considerable amount of precipitates also within the grains in Fe8. At both locations, between and within the grains, the precipitates exhibit an ellipsoidal shape with the long axis aligned along the grain boundaries in the former and in 45° angles in the latter case. The precipitates have an average length of ca. 3 µm in Fe6 and ca. 4 µm in Fe8, while the width is reduced to about 1.5 µm in Fe6 and Fe8, respectively.

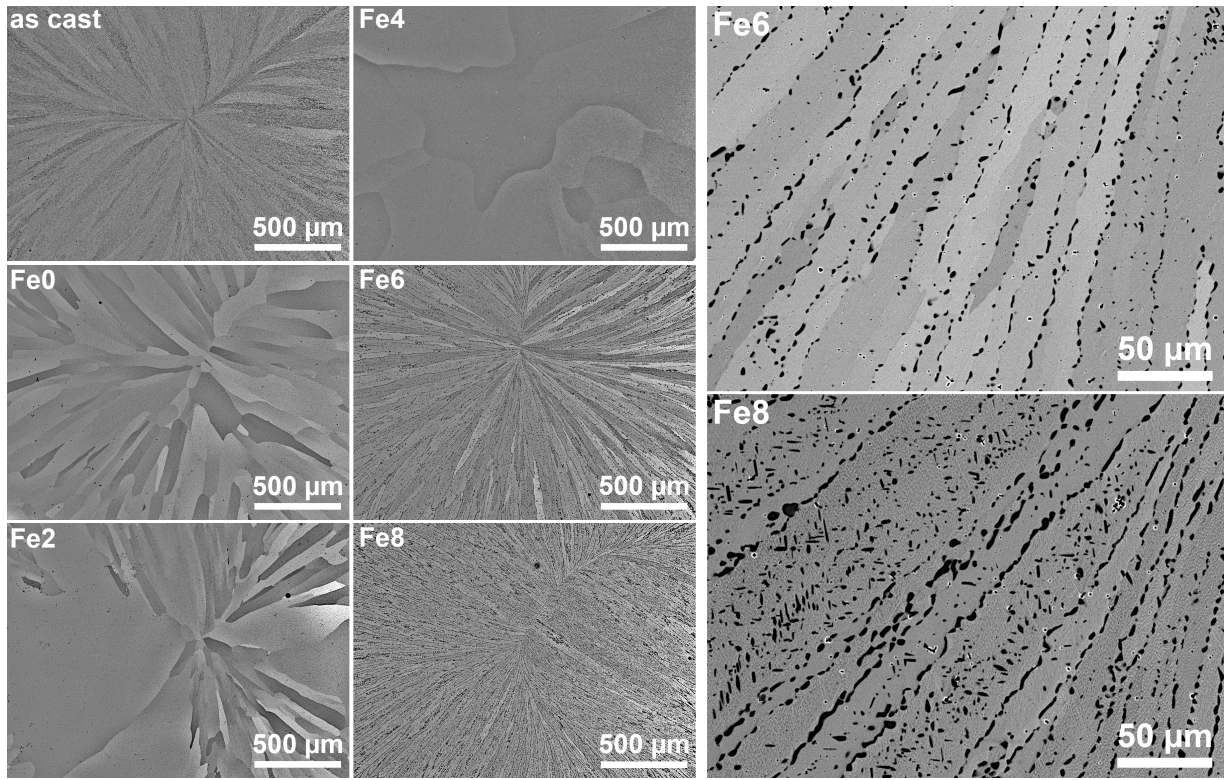

*Figure 1: BSE images of as-cast Fe0 and annealed Fe0-Fe8 samples. On the right-hand side, additional BSE images in higher resolution are provided for Fe6 and Fe8. The grey contrast corresponds to matrix and black contrast to secondary γ-phase.*

In Figure 2 room-temperature XRD patterns of the complete annealed series (Fe0 to Fe8) are displayed. Independent of the Fe-concentration, the matrix phase exhibits an austenitic $L2_1$ structure, which is indicated by the presence of the (111) reflection at ca. 12°. The diffractograms of Fe6 and Fe8 confirm the occurrence of the secondary phase, which can be indexed with a fcc A1 structure. The lattice parameter of the secondary γ-phase (a=0.3635 nm for Fe8) and the austenite matrix (a=0.5986 nm for Fe8) are in good accordance with literature data of similar stoichiometries [45]. An overview of the lattice parameters of all samples is given in Table 2.

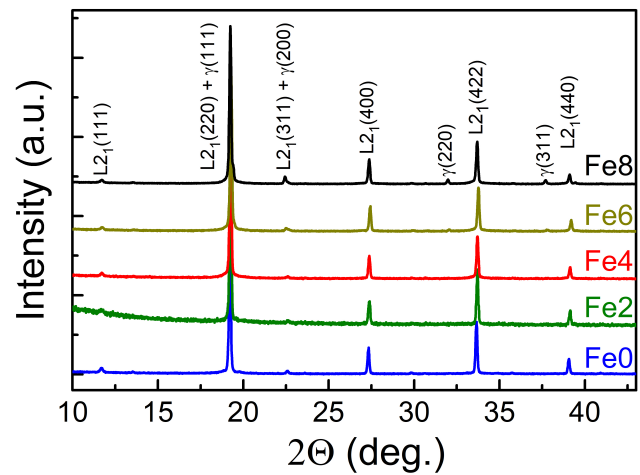

*Figure 2: XRD patterns of the annealed Fe0-Fe8 powder.*

Table 2: Lattice constants of L2$_1$ austenite matrix and A1 secondary γ-phase obtained from XRD measurements of Fe0-Fe8 powder

| Sample ID | Lattice constant (nm) | |
|---|---|---|
| | Matrix | Secondary phase |
| Fe0 | 0.5993 | |
| Fe2 | 0.5984 | |
| Fe4 | 0.5984 | |
| Fe6 | 0.5973 | 0.3626 |
| Fe8 | 0.5986 | 0.3635 |

Figure 3(a) exhibits a bright field TEM image of Fe8 in which both phases, L2$_1$ matrix and the γ-precipitates, are present. Thereby, the L2$_1$ matrix is under zone axis condition. The corresponding selected area diffraction patterns of the L2$_1$ matrix phase along the [111] zone axis (see Figure 3(b)) and the secondary γ-phase along the [110] zone axis (see Figure 3(c)) display a good agreement with the crystal structures obtained by XRD. Interestingly, the high resolution TEM images in Figure 3(d) and (e) show a similar d-spacing of the lattice planes perpendicular to the interface for both phases indicating a coherent phase boundary. *Chluba et al.* [46] have exemplarily shown for the shape-memory alloy TiNiCu that the formation of coherent precipitates can significantly improve reversibility and cyclic stability during superelastic cycling. It is also worth mentioning that in contrast to the work of *Sepehri-Amin et al.* [47] no nanoprecipitates are observed in the matrix.

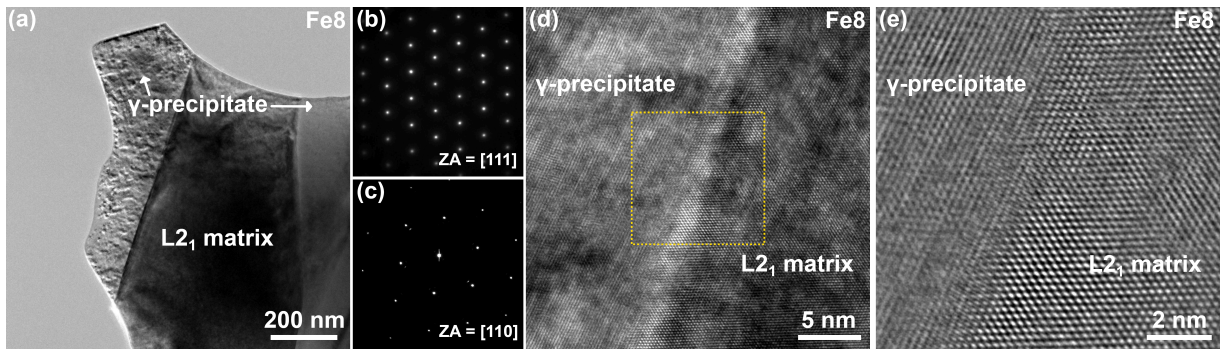

Figure 3: TEM images of Fe8. (a) Bright field TEM image. (b),(c) Selected area electron diffraction patterns of L2$_1$ matrix phase (b) and secondary γ-phase (c). (d) High-resolution TEM (HRTEM) image at the interface of L2$_1$ matrix phase and secondary γ-phase. (e) Corresponding average background subtraction filtered HRTEM image of region of interest (marked by yellow box in (d)).

3.2. Martensitic transformation behavior

3.2.1. Transition temperatures and entropy changes

In order to study the influence of Fe-doping and the corresponding microstructural changes on the martensitic transformation behavior, we performed DSC measurements which are shown in Figure 4. The exothermic peaks upon cooling and the endothermic peaks during heating can be attributed to the martensitic and its reverse transformations, respectively. The corresponding transition temperatures $T_{t,M} = (M_s+M_f)/2$ and $T_{t,A} = (A_s+A_f)/2$, thermal hysteresis $\Delta T_{hys} = T_{t,A}-T_{t,M}$ and transition entropy changes $\Delta s_t$ are summarized in the insets of Figure 4 (see also Table 3). The latter have been obtained by integrating the baseline subtracted heat flow per mass unit $\dot{Q}$ of the calorimetric peaks according to

$$\Delta s_t = \int \frac{1}{T}(\dot{Q} - \dot{Q}_{baseline})\left(\frac{dT}{dt}\right)^{-1} dT. \qquad \text{Equation 1}$$

The top left inset of Figure 4(a) exhibits a good agreement of $T_{t,M}$ and $T_{t,A}$ in Fe0-Fe8. While $T_{t,M}$ shows a variation of 14.7 K, $T_{t,A}$ differs only by 11 K. In addition, a significant increase of thermal hysteresis $\Delta T_{hys}$ can be noticed in Fe8 as compared to Fe0-Fe6, which will be discussed in detail in section 3.2.2. The good coincidence of $T_{t,M}$ and $T_{t,A}$ in Fe0-Fe8 has been achieved by adjusting the Mn/In-ratio as shown in Table 1. It should be emphasized, that a good agreement of the transition temperatures $T_t$ with respect to the Curie temperature of the austenite $T_C^A$ is crucial for a comparison of the caloric properties in metamagnetic shape-memory alloys. Gottschall et al. [48] have demonstrated exemplarily for Ni-Mn-In-(Co) that $\Delta s_t$ decreases upon increasing $T_C^A - T_t$ due to a rising magnetic entropy contribution $\Delta s_{mag}$ which counteracts the lattice entropy change $\Delta s_{lat}$ according to

$$\Delta s_t = |\Delta s_{lat}| - |\Delta s_{mag}(T)|. \qquad \text{Equation 2}$$

Within the Fe-doping series, $T_C^A$ varies only by 3.8 K. As a consequence, from Fe0 to Fe6 an almost constant $|\Delta s_t|$ of about 21.5 Jkg$^{-1}$K$^{-1}$ and 23 Jkg$^{-1}$K$^{-1}$ can be observed upon the forward and reverse martensitic transformation, respectively (see bottom right inset of Figure 4). However, a further increase of the Fe-concentration results in a significant drop of $|\Delta s_t|$ to 11.2 Jkg$^{-1}$K$^{-1}$ and 15.2 Jkg$^{-1}$K$^{-1}$ in Fe8. The decrease can be attributed to the rising phase fraction (11.5 wt%) of γ-precipitates which do not undergo a martensitic transformation. Recall that in Fe6 the secondary phase only accounts for 5.2 wt%.

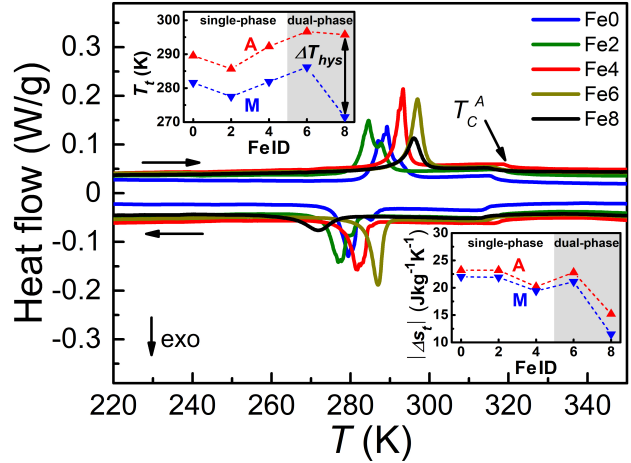

*Figure 4: DSC curves of Fe0-Fe8. The corresponding transition temperatures $T_{t,M}$, $T_{t,A}$ and transition entropy changes $\Delta s_{t,M}$, $\Delta s_{t,A}$ of Fe0-Fe8 are shown in the top left and bottom right inset, respectively. The white and gray areas in the insets indicate the single- and dual-phase samples, respectively.*

*Table 3: Martensite and austenite start and finish temperatures (i.e. $M_s$, $M_f$, $A_s$, $A_f$), corresponding transition temperatures $T_{t,M}$, $T_{t,A}$, thermal hysteresis $\Delta T_{hys}$, transition entropy changes $\Delta s_{t,M}$, $\Delta s_{t,A}$, Curie temperatures of the austenite $T_C^A$ and $\lambda_2$ eigenvalues of Fe0-Fe8. $\lambda_2$ eigenvalues were calculated from temperature-dependent XRD data (see supplementary material). All other parameters were determined from DSC measurements.*

| Sample ID | $M_s$ (K) | $M_f$ (K) | $A_s$ (K) | $A_f$ (K) | $T_{t,M}$ (K) | $T_{t,A}$ (K) | $\Delta T_{hys}$ (K) | $\Delta s_{t,M}$ (Jkg$^{-1}$K$^{-1}$) | $\Delta s_{t,A}$ (Jkg$^{-1}$K$^{-1}$) | $T_C^A$ (K) | $\lambda_2$ |
|---|---|---|---|---|---|---|---|---|---|---|---|
| Fe0 | 287.4 | 275.8 | 283.5 | 295.6 | 281.6 | 289.6 | 8 | -22 | 23.2 | 315.4 | 1.0046 |
| Fe2 | 282.0 | 273.0 | 281.0 | 290.4 | 277.5 | 285.7 | 8.2 | -21.9 | 23.2 | 318.1 | 1.0022 |
| Fe4 | 284.8 | 278.9 | 290.1 | 294.5 | 281.9 | 292.3 | 10.4 | -19.4 | 20.2 | 318.6 | 1.0025 |
| Fe6 | 288.8 | 283.6 | 294.2 | 299.1 | 286.2 | 296.7 | 10.5 | -21.1 | 22.8 | 316.4 | 1.0019 |
| Fe8 | 277.7 | 265.2 | 292.3 | 299.2 | 271.5 | 295.8 | 24.3 | -11.5 | 15.2 | 314.8 | 1.0029 |

### 3.2.2. Thermal hysteresis

A notable difference between Fe0-Fe6 and Fe8 can also be observed in the thermal hysteresis $\Delta T_{hys}$ of the martensitic transformation (see top left inset of Figure 4). While from Fe0 to Fe6 $\Delta T_{hys}$ varies between 8 K and 10.5 K only, an increase to 24.3 K can be observed for Fe8. Intrinsically, $\Delta T_{hys}$ in metamagnetic shape-memory alloys is dominated by the geometrical compatibility of the martensite and austenite state which can be estimated via the middle eigenvalue $\lambda_2$ of the transformation stretch matrix. For a rising deviation of $\lambda_2$ from unity $|\lambda_2-1|$ several studies [49–51] demonstrate a drastic increase of $\Delta T_{hys}$. The transformation stretch matrix and the corresponding eigenvalues can be obtained from the lattice parameters of martensite and austenite according to the procedure presented in [52,53]. In this work, the lattice parameters of both phases were determined by temperature-dependent XRD which is provided in detail as supplementary material. The calculated $\lambda_2$ eigenvalues are listed in Table 3. The minor deviations of $\lambda_2$ from unity indicate an excellent geometric compatibility of the matrix phase in all samples. Thereby, the compatibility in the Fe-containing samples seems to be slightly improved in comparison with Fe0. However, overall no clear dependence of $|\lambda_2-1|$ on the Fe-concentration and only a small variation of $\lambda_2$ can be found. As a consequence, no correlation of $\lambda_2$ with $\Delta T_{hys}$ is observed.

Extrinsically, microstructural aspects such as grain size or secondary phases are the main factors influencing $\Delta T_{hys}$ [47,54]. In order to explain the significantly enlarged $\Delta T_{hys}$ in Fe8, it is worth comparing the microstructures of Fe6 ($\Delta T_{hys}$ = 10.5 K) and Fe8 ($\Delta T_{hys}$ = 24.3 K). Both samples exhibit similar grain sizes but different volume fractions of γ-precipitates. While in Fe6 the precipitates occur almost exclusively at grain boundaries, a significant amount of precipitates can also be observed within the grains in Fe8. In consequence, the martensite front can easily propagate through the grains in Fe6 as shown by temperature-dependent optical microscopy in Figure 5. It should be noticed that the martensite preferentially nucleates within the grains and grows highly anisotropic, parallel to the long axis of the columnar grains (videos of the martensitic transformation in Fe6 and Fe8 are provided as supplementary material). As the growth of martensite takes place in the Heusler matrix without hindrance by the precipitates, $\Delta T_{hys}$ of dual-phase Fe6 is similar as compared to the single-phase samples. In Fe8 the increased amount of precipitates within the grains constraints the nucleation and growth of martensite, which is indicated by smaller variant sizes (see Figure 5). Hence, increased frictional work dissipation is expected causing hysteresis growth. In addition, the significantly enlarged temperature-induced transition width ($M_s$-$M_f$) in Fe8 suggests an increase of the stored elastic strain energy. *Hamilton et al.* [55] proposed that an enhanced elastic energy relaxation can be promoted by higher precipitate volume fractions resulting as well in a rising $\Delta T_{hys}$.

$T > M_s$

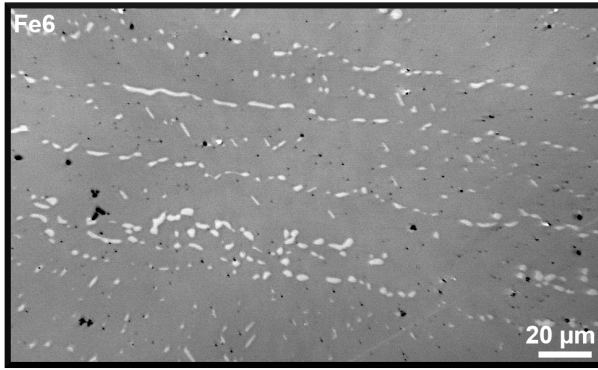 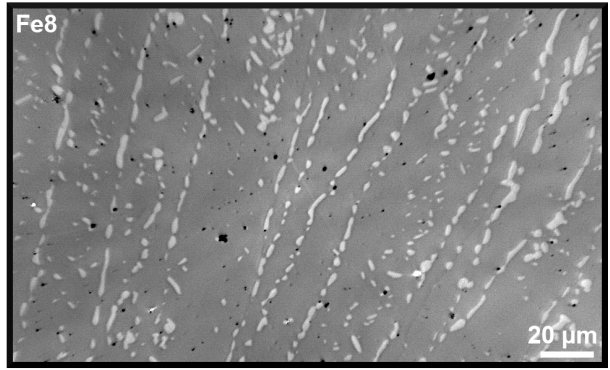

$T \approx M_s$

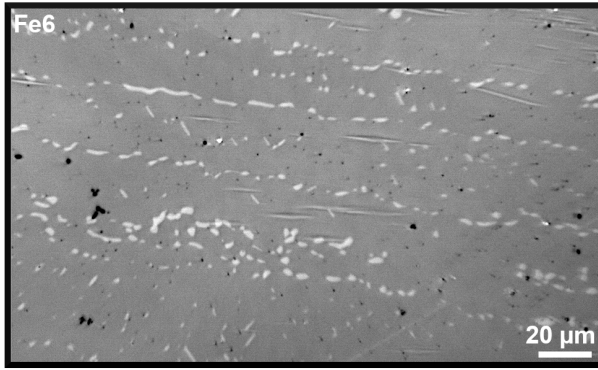 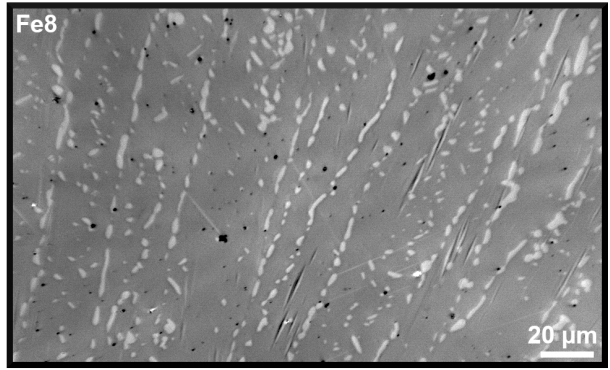

$M_s < T < M_f$

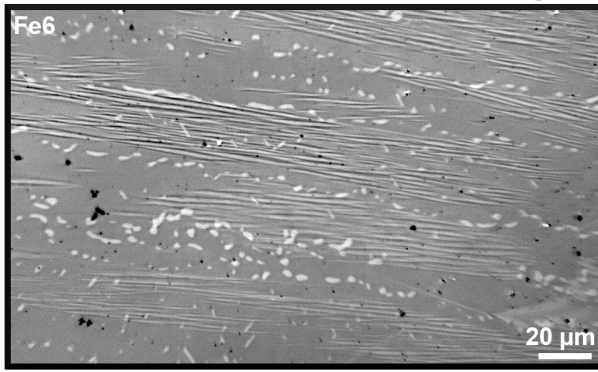 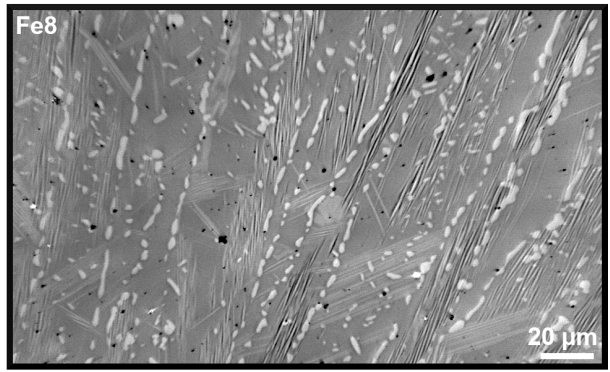

$T < M_f$

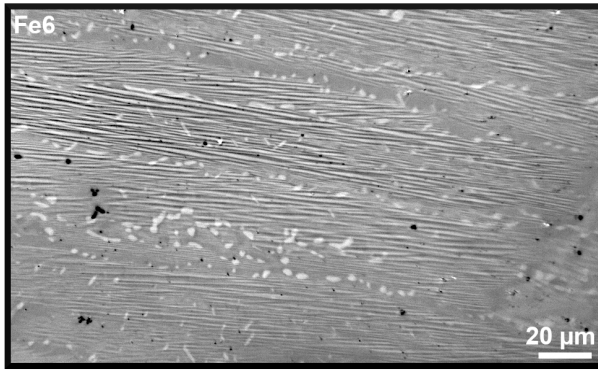 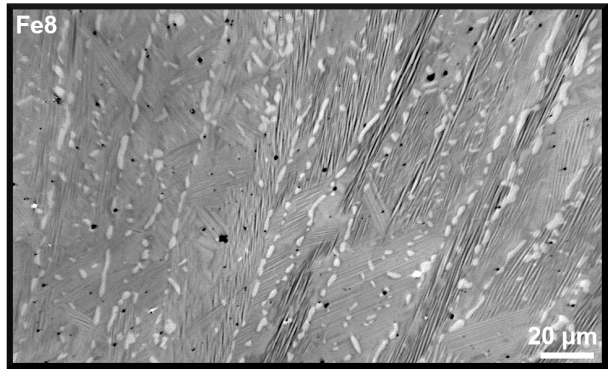

*Figure 5: Temperature-dependent optical microscopy of Fe6 (left) and Fe8 (right) upon cooling through martensitic transformation. The grey phase corresponds to matrix and the bright phase to γ-precipitates. The small black features correspond to pores and manganese oxides.*

### 3.3. Magnetocaloric effect

In order to utilize the transition entropy changes $\Delta s_t$ reported in section 3.2.1 by the application of magnetic fields, large changes of magnetization $\Delta M$ are essential. According to the Clausius-Clapeyron equation

$$\frac{\Delta M}{\Delta s_t} = -\frac{dT_t}{\mu_0 dH} \qquad \text{Equation 3}$$

high values of $\Delta M$ ensure a significant field sensitivity ($dT_t/\mu_0 dH$) of the first-order transformation and thus drive the magnetocaloric effect [56]. Figure 6 displays the isofield $M(T)$ curves in 1 T. A distinct $\Delta M$ can be observed upon the transition between low magnetization martensite and high magnetization austenite in all Fe-doped samples. In comparison with Fe0, a slight increase of the austenite magnetization is apparent for Fe2 and Fe4. With further increase of the Fe-concentration, the austenite magnetization is reduced due to the precipitation of the secondary γ-phase in Fe6 and Fe8. According to the ternary Ni-Mn-Fe magnetic phase diagram, the secondary phase is paramagnetic in the given composition and temperature range [57]. In the martensite state, Fe4, Fe6 and Fe8 exhibit an increased magnetization compared with Fe0 and Fe2 which can be attributed to compositional changes of the matrix phase. In consequence, Fe2 shows the highest and Fe8 the lowest $\Delta M$ upon the first-order transformation.

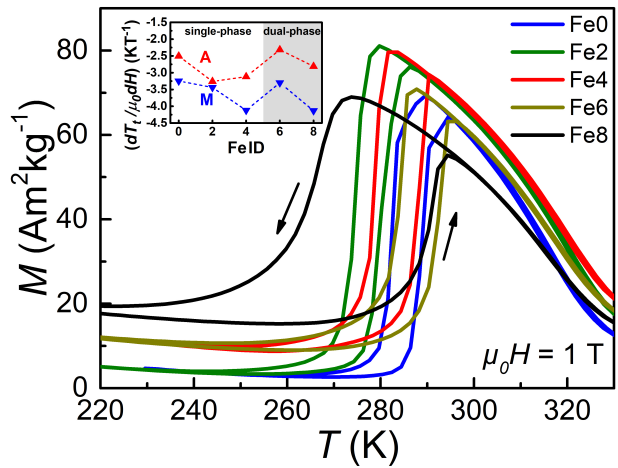

Figure 6: Isofield magnetization curves in 1 T of Fe0-Fe8. The inset depicts the field sensitivities $dT_{t,M}/\mu_0 dH$ (M) and $dT_{t,A}/\mu_0 dH$ (A) as a function of the nominal Fe content. The white and gray area in the inset indicate the single- and dual-phase samples, respectively.

The inset of Figure 6 illustrates the resulting field sensitivities of the forward ($dT_{t,M}/\mu_0 dH$) and the reverse ($dT_{t,A}/\mu_0 dH$) martensitic transformation which have been determined from $M(T)$ measurements in 0.1, 1, and 2 T by linear regression. While ($dT_{t,M}/\mu_0 dH$) varies within -3.2 KT$^{-1}$ and -4.1 KT$^{-1}$, ($dT_{t,A}/\mu_0 dH$) ranges between -2.5 KT$^{-1}$ and -3.3 KT$^{-1}$ in the Fe-doping series. The field sensitivities are in excellent agreement with the theoretical values obtained from $\Delta M/\Delta s_t$ according to equation 3. The minus-signs of ($dT_t/\mu_0 dH$) refer to a stabilization of the ferromagnetic austenite towards lower temperatures in rising magnetic fields.

The isothermal entropy changes $\Delta s_T$ (see Figure 7(a)) upon the corresponding metamagnetic transition from martensite to austenite have been computed from $M(H)$ curves (not shown here) up to 2 T using

$$\Delta s_T(T,H) = \mu_0 \int_0^H \left(\frac{\partial M(T,H)}{\partial T}\right)_H dH. \qquad \text{Equation 4}$$

For Fe0-Fe6 comparable $\Delta s_T$ values between 15 Jkg$^{-1}$K$^{-1}$ and 17 Jkg$^{-1}$K$^{-1}$ can be observed due to similar field sensitivities ($dT_{t,A}/\mu_0 dH$) and temperature-induced transition widths. In Fe8, $\Delta s_T$ decreases to 9 Jkg$^{-1}$K$^{-1}$ which can be attributed to a rising fraction of the non-transforming secondary phase as well as a broadening of the transformation near $A_s$. The corresponding adiabatic temperature changes $\Delta T_{ad}$ in field changes of 1.9 T are shown in Figure 7(b). While $|\Delta T_{ad}|$ varies within 3.0 K and 3.7 K for Fe0-Fe6, a decrease to 2.5 K can be noticed in Fe8. Compared to $\Delta s_T$, the decay in Fe8 is less pronounced as $\Delta T_{ad}$ is an intensive variable and with this not directly reduced by the secondary phase acting as inactive mass. In this case, the slight reduction of $|\Delta T_{ad}|$ in Fe8 can result from the aforementioned transition broadening and the heat transfer between matrix and secondary phase. It is worth mentioning that despite the reduction of $\Delta s_T$ and $|\Delta T_{ad}|$ in Fe8, a good agreement with literature values for "Fe-free" Ni-Mn-In of similar transition temperatures and field changes is obtained [47,58–60].

The peak-like shape of $\Delta s_T$ and $\Delta T_{ad}$ in all samples indicates that the reverse martensitic transformation is not completed in field changes of 2 T and 1.9 T, respectively. This is also confirmed by comparing $\Delta s_T$ with the transition entropy changes $\Delta s_{t,A}$ determined from DSC (see Table 3). Hence, even higher values of $\Delta s_T$ and $\Delta T_{ad}$ can be achieved upon the reverse martensitic transformation when the magnetic field change is raised further [61]. The negative (positive) values of $\Delta s_T$ ($\Delta T_{ad}$) at high temperatures result from the conventional magnetocaloric effect being present around $T_C^A$.

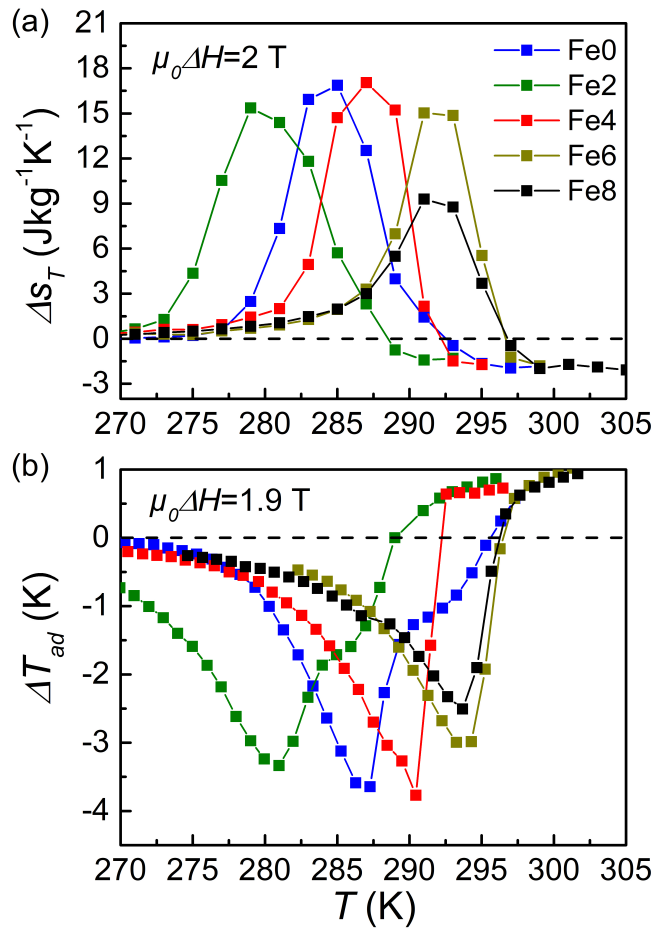

Figure 7: (a) Isothermal entropy changes $\Delta s_T$ of reverse martensitic transformation in magnetic field changes of 2 T and (b) adiabatic temperature changes $\Delta T_{ad}$ in magnetic field changes of 1.9 T of Fe0-Fe8.

### 3.4. Mechanical properties

Figure 8(a) displays the compressive stress-strain curves of Fe0-Fe8 at 300 K. Hence, all samples are initially in the austenite state. At small strains, the linear elastic region of the austenite is followed by the onset of a stress-induced martensitic transformation at $\sigma^{Ms}$. Compared to Fe0 and Fe2, a minor increase of $\sigma^{Ms}$ can be observed for Fe4-Fe8. Upon the transition, increasing slopes are apparent for higher Fe-concentrations. In Fe6 and Fe8, this can be attributed to the presence of γ-precipitates hindering the martensite growth by internal stress fields or dislocation generation at the matrix/precipitate phase boundary [62,63]. *Xu et al.* [64] report similar effects for the precipitation of γ-phase at grain boundaries in Ni-Fe-Ga magnetic shape-memory alloys. It should be emphasized that the effect of precipitates on the stress-induced martensitic transformation can significantly differ from the thermally-induced transition described in section 3.2.2. The rising transformation slopes in Fe2 and Fe4 could be attributed to changes in grain orientation/texture due to abnormal grain growth [34].

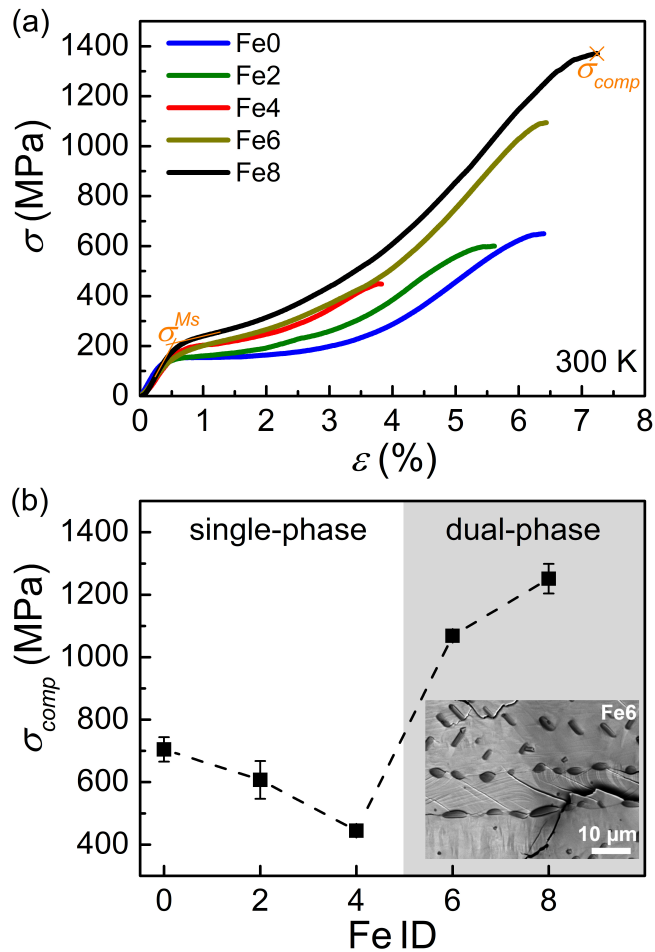

*Figure 8: (a) Compressive stress-strain curves of Fe0-Fe8 measured at 300 K. The martensite start stress $\sigma^{Ms}$ and the compressive strength $\sigma_{comp}$ are exemplarily indicated for Fe8. (b) Compressive strength $\sigma_{comp}$ as a function of the Fe-content. The white and gray area indicate the single- and dual-phase samples, respectively. The inset shows the fracture surface of Fe6.*

When the strain is increased further, elastic deformation and subsequently yielding of the martensite can be noticed until fracture occurs at the compressive strength $\sigma_{comp}$. Average values of $\sigma_{comp}$ were obtained by testing at least two specimens per Fe-content and are summarized in Figure 8(b). While Fe0 exhibits a $\sigma_{comp}$ of about 700 MPa, a decrease to 400 MPa can be noticed in Fe4. This decay can result from the significantly enlarged grains in Fe4. In contrast to that, an increase of $\sigma_{comp}$ to 1068 MPa and 1233 MPa is apparent for Fe6 and Fe8, respectively. This can be attributed to the presence of precipitates hindering intergranular fracture which is typically observed in single-phase Ni-Mn-based Heusler compounds [65–67]. As shown in the inset of Figure 8(b), the γ-precipitates lead to crack deflection at grain boundaries. In addition, dislocation loops can be observed in the direct surrounding of the precipitates. Besides the toughening caused by precipitates, smaller grain sizes in Fe6 and Fe8 can further enhance $σ_{comp}$.

### 3.5. Elastocaloric effect

#### 3.5.1. Stress-dependence of the elastocaloric effect

The influence of Fe-doping on the elastocaloric performance was investigated by monitoring the stress-induced adiabatic temperature change $ΔT_{ad}$ upon compressive superelastic cycling up to various stresses at a starting temperature of $A_f$ + 20 K. Adiabatic conditions were simulated by a strain rate of $3×10^{-2}$ s$^{-1}$ during loading and unloading [44]. In addition, a holding time of 25 s was introduced for thermal relaxation. The resulting temperature-time profiles and the corresponding stress-strain curves are exemplarily shown for Fe6 in Figure 9(a). The increasing temperature at about 6 s can be attributed to the stress-induced martensitic transformation during loading and the decreasing temperatures at about 20 s refer to its reverse transformation upon unloading. A rising |$ΔT_{ad}$| is apparent at higher maximum stresses. While for application and removal of 200 MPa, a $ΔT_{ad}$ of +1.9 K and -2.6 K can be observed, a stress of 600 MPa results in a $ΔT_{ad}$ of +9.5 K and -9 K, respectively. This originates from an increasing transformed phase fraction at higher stresses which is indicated by rising superelastic strains and stress hysteresis in the inset of Figure 9(a). Accordingly, a saturation of $ΔT_{ad}$ can be observed when the stress is high enough to fully induce the transformation.

Figure 9(b) depicts the maximum values of $ΔT_{ad}$ upon loading and unloading as a function of applied stress for Fe0-Fe8. Missing data points for Fe0-Fe4 at high stresses are caused by material failure during the experiment due to a lower compressive strength of these alloys. It is found that the stress required to saturate $ΔT_{ad}$ increases with higher Fe-concentrations. While in Fe0 and Fe2 about 300 MPa are sufficient, ca. 500 MPa are required in Fe8. This can be explained by the increasing transition slopes at higher Fe-concentrations (see Figure 8(a)).

To evaluate the influence of Fe-doping on the saturated elastocaloric effect, the highest values of |$ΔT_{ad}$| in Fe0-Fe8 are compared. Upon unloading, similar $ΔT_{ad}$ values of about -9 K are obtained for Fe0-Fe6 as the transition entropy changes of the matrix phase are in good agreement. As discussed in section 3.2.1, the transition entropy change of the matrix is significantly lowered in Fe8 resulting in a $ΔT_{ad}$ of about -6 K upon unloading. Additionally, the heat transfer from matrix to secondary phase can contribute to the decrease of |$ΔT_{ad}$| in Fe8. During loading, a similar dependence of the saturated elastocaloric effect on the Fe-concentration is found. Thereby, about 0.5-2 K higher values of |$ΔT_{ad}$| are obtained as compared to unloading due to frictional heating [68].

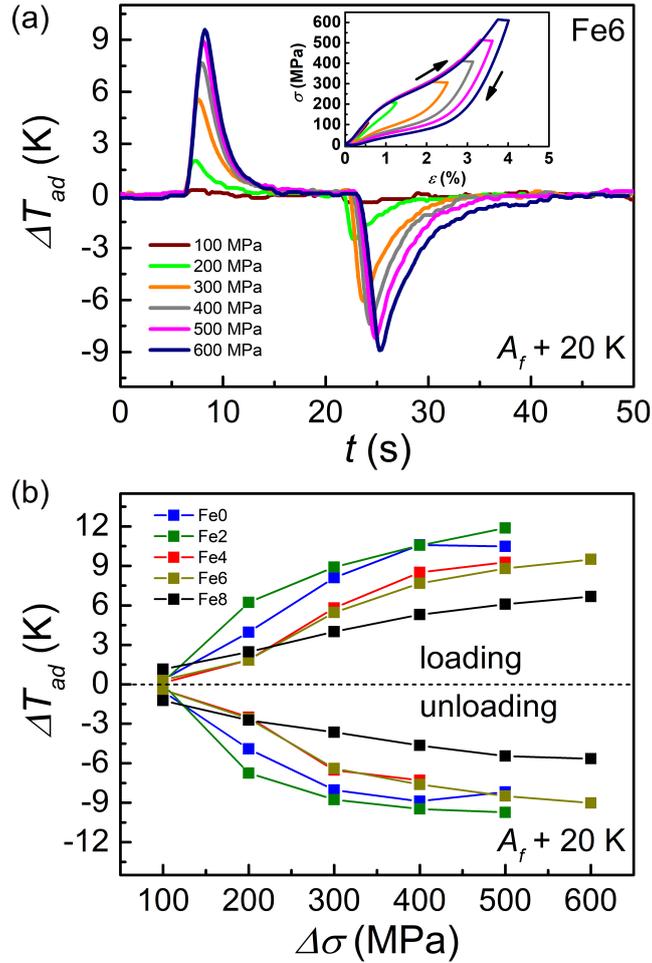

*Figure 9: (a) Temperature-time profile of Fe6 during compressive superelastic cycles up to various stresses. In the inset the corresponding stress-strain curves are shown. (b) Adiabatic temperature change ΔT$_{ad}$ as function of the applied compressive stress (loading and unloading) in Fe0-Fe8. All samples were tested at a starting temperature of A$_f$ + 20 K.*

### 3.5.2. Cyclic stability

In order to analyze the influence of Fe-doping on the cyclic stability, temperature-time profiles were recorded upon compressive superelastic cycling with a maximum stress of 300 MPa. The starting temperature of $A_f$ + 20 K and the strain rate of $3\times10^{-2}$ s$^{-1}$ during loading and unloading are equal to the measurements presented in the previous section. Solely, the holding time after loading was reduced to 15 s due to time considerations. In Figure 10(a) and (b) the temperature-time profiles are exemplarily depicted for Fe0 and Fe6. Figure 10(c) summarizes the number of cycles and the cyclic $\Delta T_{ad}$ upon unloading for Fe0-Fe8. In Fe0, a cyclic $\Delta T_{ad}$ of -8 K upon unloading is found. However, after 10 cycles degradation is visible until brittle fracture occurred after 15 cycles. When the Fe-concentration is increased, from Fe0 to Fe4, the cycles to failure slightly decrease as a result of the grain growth. Note that a similar trend can be observed for the compressive strength (see Figure 8(b)). Accordingly, a drastic increase of the cyclability is found for Fe6 and Fe8. Both samples were cycled more than 16000 times without functional or structural degradation. This corresponds to an increase of more than three orders of magnitude as compared to Fe0 which can be attributed to the formation of the secondary γ-phase hindering crack propagation upon cyclic loading. Thereby, a cyclic $\Delta T_{ad}$ of -4.5 K in Fe6 and -3.2 K in Fe8 is observed.

As discussed earlier, the reduction of $\Delta T_{ad}$ as compared to Fe0 and Fe2 is caused by a smaller transformed phase fraction at the maximum stress of 300 MPa in Fe6 and Fe8. To account for this effect, additional cycling experiments were carried out with a maximum stress of 200 MPa for Fe0 and Fe2. In this case, the cyclic values of $\Delta T_{ad}$ are similar to the one in Fe6 indicating similar transformed phase fractions. Though the number of cycles increases to 194 in Fe0 and 42 in Fe2 a significant discrepancy to Fe6 and Fe8 is found.

The superior performance of Fe6 and Fe8 is also outlined when the number of cycles and the corresponding cyclic $\Delta T_{ad}$ upon unloading are compared to other metamagnetic shape-memory materials such as Ni-Mn-Ga- [69–72] , Ni-Mn-In- [30,31,73–77], Ni-Mn-Sn- [67,78,79], Ni-Mn-Ti- [80] and Ni-Mn-Al-based [81] alloys (see Figure 10(d)). While the cyclic elastocaloric effect is maintained, the cyclability is increased by almost one order of magnitude.

It should be emphasized, that the combination of a high cyclic stability with a large elastocaloric and magnetocaloric effect, as achieved for Fe6, excellently meets the requirements for multicaloric cooling applications in which magnetic fields and uniaxial are applied sequentially [23] or simultaneously [24]. This way, even larger cyclic caloric effects can be achieved as compared to their single caloric counterparts.

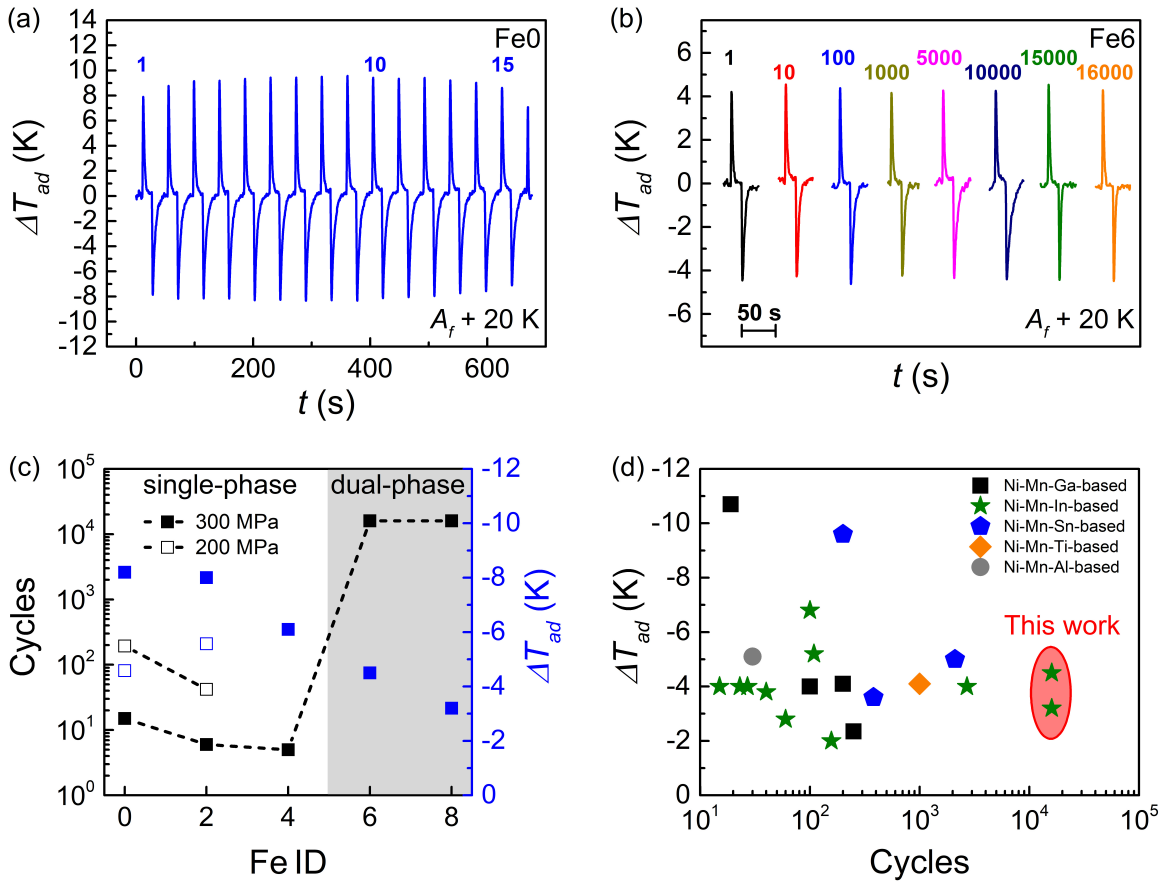

Figure 10: (a), (b) Temperature-time profiles of Fe0 and Fe6 during compressive cyclic loading up to 300 MPa. (c) Number of cycles and corresponding cyclic $\Delta T_{ad}$ of Fe0-Fe8 for compressive cyclic loading up to 300 MPa. Additional experiments were carried out up to 200 MPa for Fe0 and Fe2. All measurements were performed at a starting temperature of $A_f$ + 20 K. For Fe0-Fe4, the number of cycles corresponds to the cycles to failure. The white and gray area indicate the single- and dual-phase samples, respectively. (d) Comparison of cyclic $\Delta T_{ad}$ and number of cycles achieved in this work (Fe6 and Fe8) with other metamagnetic shape-memory materials (Ni-Mn-Ga- [69–72] , Ni-Mn-In- [30,31,73–77], Ni-Mn-Sn- [67,78,79], Ni-Mn-Ti- [80] and Ni-Mn-Al-based [81] alloys) upon cyclic loading. The number of cycles corresponds to the maximum values reported in the literature independent on the occurrence of failure. The cyclic $\Delta T_{ad}$ in (c) and (d) corresponds to the values upon unloading.

## 4 Conclusions

This work presents a microstructure design strategy to simultaneously achieve large magnetocaloric and elastocaloric effects with superior cyclic stability in Ni-Mn-In Heusler compounds by alloying with Fe. We demonstrate that exceeding a solubility limit of 4.5 at% Fe in the Ni-Mn-In Heusler matrix results in precipitation of a strongly Fe-rich and In-depleted γ-phase during annealing. The γ-phase forms a coherent interface with the Heusler phase matrix and is preferentially nucleated at grain boundaries hindering intergranular fracture during cyclic loading. As a result, a large elastocaloric effect of -4.5 K could be observed in >16000 cycles without degradation. This corresponds to an increase of the cyclic stability by more than three orders of magnitude as compared to single-phase Ni-Mn-In-(Fe). If Ni-Mn-In is Fe-doped within the solubility limit, grain coarsening occurs which reduces the cyclic stability.

At the same time, the excellent magnetocaloric properties of single-phase Ni-Mn-In-(Fe) can be maintained in dual-phase samples when the amount of Fe is adjusted in a way that precipitates are almost exclusively present at grain boundaries. In this case, the martensite can propagate in the Heusler matrix without hindrance by the γ-phase resulting in a similar thermal hysteresis and transition width. In combination with a large change in magnetization, an isothermal entropy change of 15 Jkg$^{-1}$K$^{-1}$ and an adiabatic temperature change of -3 K have been obtained in a magnetic field of 2 T, which can be provided by permanent magnet assemblies. When the amount of Fe is increased further, and with this a considerable amount of γ-phase is formed within the grains of the Heusler matrix, the transition width and thermal hysteresis drastically increase causing a degradation of the magnetocaloric effect.

Besides a significant improvement of the cyclic stability in single-caloric cooling applications, our new microstructure design provides excellent suitability for multicaloric cooling using magnetic field and uniaxial stress. By a suitable sequential [23] or simultaneous [24] combination of both stimuli cyclic caloric effects are likely to be raised even further.

### Acknowledgements

We thank the European Research Council (ERC) under the European Union's Horizon 2020 research and innovation programme (grant no. 743116—project Cool Innov) and the Deutsche Forschungsgemeinschaft (DFG, German Research Foundation) – Project-ID 405553726 – TRR 270 for funding this work.